\documentstyle[prl,aps,multicol,psfig]{revtex}

\begin{document}

\title{Suppression of carrier induced ferromagnetism by composition and
spin fluctuations in diluted magnetic semiconductors}
\author{Yu.G.Semenov}
\address{Groupe d'Etude des Semi-Conducteurs, UMR 5650 CNRS-Universit\'{e}
Montpellier 2,\\
Place Eug\`{e}ne Bataillon, 34095 Montpellier Cedex, France}
%\address{Institute of Physics of Semiconductors, NASC of Ukraine, Prosp.\ Nauki 45,\\
%252650, Kiev, Ukraine}
\author{V.A.Stephanovich}
\address{Institute of Mathematics, University of Opole, Oleska 48, 45-052, Opole, Poland}

\date{\today}
\maketitle
\begin{abstract}

We suggest an approach to account for spatial (composition) and thermal fluctuations
in "disordered" magnetic models (e.g. Heisenberg, Ising) with given spatial
dependence of magnetic spin-spin interaction. Our approach is based on introduction
of fluctuating molecular field (rather than mean field) acting between the spins.
The distribution function of the above field is derived self-consistently. In general
case this function is not Gaussian, latter asymptotics occurs only at sufficiently
large spins (magnetic ions) concentrations $n_i$.
Our approach permits to derive the equation for a critical temperature $T_c$ of
ferromagnetic phase transition with respect to the above fluctuations.
We apply our theory to the analysis of influence of composition
fluctuations  on $T_c$ in diluted magnetic semiconductors (DMS) with RKKY
indirect spin-spin interaction.
%We compare our theory with mean field approximation
%as well as with Gaussian asymptotics for the distribution function.
Two new results are obtained. (i)
It is shown that composition fluctuations destroy ferromagnetic ordering in DMS if
the carrier concentration $n_e$ exceeds the critical value $n_c\approx 0.1n_i
$.(ii) In the case $n_e<0.5n_c$ and
magnetic ion spin $S=5/2$ the composition fluctuations shift the critical
temperature $T_c$ (as compared to its mean field value)
by a factor about $1-5n_e/n_i$.
\end{abstract}

\begin{multicols}{2}

\narrowtext

Ferromagnetic (FM) ordering in p-doped diluted magnetic semiconductors (DMS)
attracts much attention of both experimentalists \cite
{Story}-\cite{Ohno} and theorists \cite{PR}-\cite{MDon}. Common approach
to this problem is a mean field approximation (MFA), corresponding to
some mean overall (homogeneous) carrier and magnetic ion
magnetization. Once appeared, the exchange fields of carriers and magnetic
ions lead to mutual spin splitting that can be stable at low enough
temperatures $T<T_c$\cite{Zeener,PR}.
The problem of "high temperature ferromagnetism" in DMS is now very important
from the point of view of technical applications. So, the question
how to increase the critical temperature $T_c$ has been widely discussed
in recent publications(see\cite{Dietl01} and references therein). One way to control
$T_c$ is to vary the concentrations of magnetic ions and/or carriers.
%%%%%%%%%%%%%%%%%%%%%%%%%%%%%%%%%%%%%%%%%%%%%%%%%%%%%%%%%%%%%%%%%%%%%%%%%%%%%%%%%%%%%%%%%%%
In the case of degenerate carriers, the critical temperature $T_c$ of FM
phase transition in the bulk DMS is proportional to $n_in_e^{1/3}$\cite{PR},
where $n_i$ and $n_e$ are the magnetic ion and carrier concentrations respectively.

%{\bf ne yasno, zachem eta fraza }The effect of magnetic ions concentration is limited by strong antiferromagnetic %(AFM)
%ion-ion interaction that blocks FM ordering, while electron
%concentration seems to be increasing without physical limitations in MFA.

Beyond the MFA, we should consider the Friedel oscillations of carrier
spin polarization\cite{KittelBook} occurring at
the scale of inverse Fermi wave vector $1/k_F$. In this case, the fluctuations of
inter-magnetic ion distance $r_{i,j}$ should be taken into account, if the
nearest neighbor mean value ${\overline r}$ is about $1/k_F$. Really, in the
case of $k_F{\overline r}\approx 1$ the numbers of
magnetic ion pairs with FM and AFM RKKY interaction is comparable, which can lead to
suppression or complete destruction of long range FM order.

Note, that the limitation of MFA by inequality $%
k_F\overline{r}<<1$ was pointed out earlier\cite{Dietl01}, but quantitative
contribution of disorder in the magnetic ion subsystem has not been considered till
present. Moreover, the problem of the existence or non-existence of a finite
critical temperature $T_c>0$ at arbitrary relation between $n_i$ and $n_e$ has
not been resolved also.
%%%%%%%%%%%%%%%%%%%%%%%%%%%%%%%%%%%%%%%%%%%%%%%

Here we develop a theory of magnetic ordering of disordered magnetic system
with given spin-spin interaction in terms of fluctuating local molecular
field approximation (FFA). The method we propose here is applied to
calculations of $T_c$ caused by RKKY interaction.
%To compare the results for $T_c$ obtained in FFA and MFA the
%numerical calculations were performed for the case of RKKY spin-spin
%interaction in DMS.

The Hamiltonian of the problem reads
\begin{equation}
{\cal H}=\sum_{j<j'}J({\vec r}_{j,j'}){\vec S} _j{\vec S} _{j'}+\sum_j%
{\vec H}_0 {\vec S}_j,  \label{mu0}
\end{equation}
where magnetic field ${\vec H}_0$ and interaction $J(r_{j,j'})$ is measured in energy units
(i.e. $g\mu =1$, $\mu$ is Bohr magneton).
%The external magnetic field
%${\vec H}_0=\{0,0,H_0\}$ is directed along $OZ$ axis.
The Hamiltonian (\ref{mu0}) incorporates
two "sources of randomness". First, (the spatial disorder) is that spin can be randomly
present or absent
in the specific $j$-th cite of a host semiconductor. Second, (the thermal disorder) is a random
projection of a spin (if any) in $j$-th cite.

Note, that our general formalism is valid
for any form of $J({\vec r}_{j,j'})$ so that its specific form will be chosen later.

In a mean field approximation, the Hamiltonian (\ref{mu0}) reduces to the sum of Zeeman
energies
\begin{equation}
{\cal H}=\sum_i({\vec H}_i+{\vec H}_0){\vec S}_i  \label{Zeem}
\end{equation}
in the external magnetic field ${\vec H}_0$ and local molecular field
\begin{equation}
{\vec H}_i=\sum_{j \neq i}J({\vec r}_{i,j})
\left\langle {\vec S} _j\right\rangle ,  \label{Hj}
\end{equation}
where $\left\langle {\vec S}_j\right\rangle $ is a thermal
average at a site $j$. Next step of the MFA is to substitute
the molecular field (\ref{Hj}) by the mean field
\begin{equation}
{\vec H}_{mf}=\overline{\sum_{j\neq i}J({\vec r}_{i,j})\left\langle {\vec S}_j
\right\rangle }.  \label{Hmf}
\end{equation}
The bar means the averaging over spatial disorder
%(i.e. all sites occupied by spins)
, so the ${\vec H}_{mf}$ is homogeneous over crystal volume; $%
\left\langle {\vec S}\right\rangle $ is an average spin in this
mean field,
\begin{equation}
\left\langle {\vec S}\right\rangle =\frac{{\rm Tr}\{{\vec S}
e^{-\beta ({\vec H}_{mf}+{\vec H}_0){\vec S}}\}}{%
{\rm Tr}e^{-\beta ({\vec H}_{mf}+{\vec H}_0){\vec S}}},
\label{Smf}
\end{equation}
$\beta =1/T$. The system of Eqs (\ref{Hmf}) and (\ref{Smf}) gives well-known
solution for mean fields ${\vec H}_{mf}$ and magnetization ${\vec M}=ng\mu
\left\langle {\vec S}\right\rangle $.

This work draws attention that the spatial and thermal fluctuations can be taken into
consideration by introduction of random field rather than mean field.
In our approach, we consider every spin ${\vec S}_j$ as a source of fluctuating (random) field
\begin{equation}
{\vec H}_f \equiv J({\vec r}_i-{\vec r}_j){\vec S}_j \label{hf}
\end{equation}
affecting other spins at the sites ${\vec r}_i$. In other words,
every spin in our approach is subjected to some random (rather then mean) field,
created by the rest of spin
ensemble. So, all thermodynamic properties of the system
will be determined by the distribution function $f({\vec H}_f)$ of the random
field ${\vec H}_f$. Namely, any spin dependent macroscopic quantity (like magnetization)
$<<A>>$ reads
\begin{equation}
<<A>>=\int <A>_{{\vec H}_f}f({\vec H}_f)d{\vec H}_f,  \label{eq2}
\end{equation}
where $<A>_{{\vec H}_f}$ is single particle thermal average with
effective Hamiltonian (\ref{Zeem}), where ${\vec H}_i$ is substituted by
${\vec H}_f$.
% single particle Zeeman energy in a
%random field ${\vec H}_f$ (rather than mean field (\ref{Hmf})and external field $H_0$:
%\begin{equation}
%<A>_{{\vec H}_f}=\frac{{\rm Tr}\{Ae^{-\beta ({\vec H}_f+{\vec H}_0){\vec %
%S}}\}}{{\rm Tr}e^{-\beta ({\vec H}_f+{\vec H}_0){\vec S}}}%
%,  \label{eq3}
%\end{equation}
%thermal
%averaging of any spin operator $A=A({\widehat S}_X,{\widehat S}_Y,{\widehat %
%S}_Z)$ should be considered with respect to one particle Zeeman energy in a
%random field ${\vec H}_f$ (rather than mean field (\ref{Hmf})and external field $H_0$:
%\begin{equation}
%<A>_{{\vec H}_f}=\frac{{\rm Tr}\{Ae^{-\beta ({\vec H}_f+{\vec H}_0){\vec %
%S}}\}}{{\rm Tr}e^{-\beta ({\vec H}_f+{\vec H}_0){\vec S}}}%
%,  \label{eq3}
%\end{equation}
%while following averaging over fluctuated magnetic field $H_f$ with
%distribution function $f({\vec H}_f)$ gives the mean value over ensemble of
%spins. We denote this kind of averaging as
%\begin{equation}
%<<A>>=\int <A>_{{\vec H}_f}f({\vec H}_f)d{\vec H}_f{.}  \label{eq2}
%\end{equation}

The distribution function  $f({\vec H}_f)$ is defined as

\begin{equation}
f({\vec H}_f)=\left\langle \overline{\delta \left( {\vec H}_f-\sum_{j(\neq
i)}J({\vec r}_i-{\vec r}_j){\vec S}_j-{\vec H}_0\right) }%
\right\rangle .  \label{fHf}
\end{equation}
Our FFA approach is based on micro-canonical
statistical theory of magnetic resonance line shape\cite{Stoneham}.
Latter theory assumes the additivity of local molecular field
contributions as well as the non-correlative
spatial distributions of spins (magnetic ions).

Latter assumptions with respect to
spectral representation of $\delta $ function
permit to transform the Eq. (%
\ref{fHf}) to the self-consistent integral equation for $f({\vec H}_f)\equiv f({\vec H})$. Introducing
the probability $n_i({\vec r)}d^3r$ for small volume $d^3r$
to be occupied by a particle, we obtain
\end{multicols}
\widetext
\noindent\rule{20.5pc}{0.1mm}\rule{0.1mm}{1.5mm}\hfill

\begin{equation}
f({\vec H})=\int \exp \left[ i{\vec \tau }({\vec H}-{\vec H}%
_0)\right] \exp \left( \int_V<<\exp \left[ -iJ({\vec r}){\vec S}%
{\vec \tau }\right] -1>>n_i({\vec r})d^3r\right) \frac{%
d^3 \tau}{(2\pi )^3},  \label{eq4}
\end{equation}

where according to definition (\ref{eq2})
\begin{equation}
<< \exp [ -iJ({\vec r}){\vec S}{\vec \tau }] -1 >> =
\int_{-\infty }^\infty
\frac{{\rm Tr} \left[ \exp \left( -iJ(r){\vec S}{\vec \tau }\right) \exp (-\beta ({\vec H}+{\vec H}_0){\vec S})\right]}
{{\rm Tr}\left[ \exp \left( -iJ(r){\vec S}{\vec \tau }\right)\right]}
f({\vec H})d^3H-1 \label{eq5}
\end{equation}

\hfill\rule[-1.5mm]{0.1mm}{1.5mm}\rule{20.5pc}{0.1mm}
\begin{multicols}{2}
\narrowtext

Eqs (\ref{eq4}) and (\ref{eq5}) represent the integral equation for
distribution function $f({\vec H})$. This equation can be solved iteratively.

However, in many cases it is possible to avoid the solution of the
integral equation since in these cases it is exactly reducible to the set of transcendental
equations for macroscopic quantities like $<<{\vec S}>>$ (magnetization),...,
$<<{\vec S}^n>>$, $n>1$ of the system.
%(this can be regarded as an equilibrium order parameter of spin glass\cite{Corenblit})

%On the other hand, the calculations of critical temperature of phase
%transition $T_c$ can be associated with foundation of lose stability
%conditions for solution of this equation. Last problem, which seams to be
%more simple, will be considered below.

This approach can be most easily demonstrated for the Ising case in Hamiltonian (\ref{mu0}).
Although this case is less general than Heisenberg one, it is perfectly valid
either for the case
of hole induced ferromagnetism in quantum wells or in the case of host
semiconductors with uniaxial symmetry.
%Next assumption consisting in Ising's approximation of spin-spin interaction
%is not necessary but makes calculation more simple for the purpose to
%compare the MFA and FFA . This approximation can be applied also in

If the magnetic fields are directed along OZ axis,
the scalar product reduces to ${\vec S}{\vec \tau }=S_Z\tau =m\tau $, $m=-S,...,S$
and (\ref{eq4}) takes the form
\begin{equation}
f(H)=\frac 1{2\pi }\int_{-\infty }^\infty e^{iH\tau }e^{{\cal G}(\tau
)}d\tau ;  \label{eq6}
\end{equation}
\begin{equation}
{\cal G}(\tau )=\left\langle \left\langle \int_Vn_i(\vec r)\left(
e^{-iJ(\vec r)m\tau }-1\right) d^3r\right\rangle \right\rangle .
\label{eq7}
\end{equation}
The calculation of average in (\ref{eq7}) yields

\end{multicols}
\widetext
\noindent\rule{20.5pc}{0.1mm}\rule{0.1mm}{1.5mm}\hfill

\begin{eqnarray}
&&\left\langle \left\langle e^{-iJ(\vec r)m\tau }-1\right\rangle \right\rangle
=\sum_{m=1/2}^S\left\{ a_m[\cos (J({\vec r})m\tau )-1]+ib_m%
\sin (J({\vec r})m\tau )\right\} ,  \label{eq11} \\
&&a_m =\int A_m(\beta H)f(H)dH,\ b_m =\int B_m(\beta H)f(H)dH. \nonumber \\
&&A_m(\beta H)=2\frac{\cosh (m\beta H)\sinh (\beta H/2)}{\sinh
((S+1/2)\beta H)}, \
B_m(\beta H)=2\frac{\sinh (m\beta H)\sinh (\beta H/2)}{\sinh
((S+1/2)\beta H)}.  \nonumber
\end{eqnarray}

\hfill\rule[-1.5mm]{0.1mm}{1.5mm}\rule{20.5pc}{0.1mm}
\begin{multicols}{2}
\narrowtext

In this case Eq.(\ref{eq7}) assumes the form
\begin{eqnarray}
%\begin{equation}
{\cal G}(\tau )&\equiv &{\cal G}(\left\{ a_m,b_m\right\} ,\tau
)=
\nonumber \\
&=&\sum_{m=1/2}^S\left\{ a_m{\cal F}_1(m\tau )+ib_m{\cal F}_2(m\tau
)\right\} ;  \label{eq13}\\
%\end{eqnarray}
%\end{equation}
%\begin{eqnarray}
%\begin{equation}
{\cal F}_1(m\tau )&=&\int _V n_i({\vec r})[\cos (J({\vec r})m\tau )-1]d^3r,
\nonumber \\
{\cal F}_2(m\tau ) &=&\int_V n_i({\vec r})\sin (J({\vec r})m\tau )d^3r.
\label{eq13a}
\end{eqnarray}
%\end{equation}

Let us return to Eq.(\ref{eq4}) for $f(H)$. Multiplying it by $%
A_m(\beta H)$ and $B_m(\beta H)$ and integrating over $H$,
we obtain the closed system
of nonlinear equations for {\em the numbers} $a_m$ and $b_m$. The explicit form
of this system reads
%\begin{eqnarray}
%a_m &=&\int_{-\infty }^\infty {\cal A}_m(\tau ,\beta )e^{{\cal G}%
%(\left\{ a_m,b_m\right\} ,\tau )}d\tau ;  \nonumber \\
%b_m &=&i\int_{-\infty }^\infty {\cal B}_m(\tau ,\beta )e^{{\cal G}%
%(\left\{ a_m,b_m\right\} ,\tau )}d\tau ,  \label{eq14}
%\end{eqnarray}
%where we introduce Fourier transformations
%\begin{eqnarray}
%{\cal  A}_m(\tau ,\beta ) &=&\frac 1\pi \int_0^\infty A_m(\beta H)\cos
%(H\tau )dH;  \nonumber \\
%{\cal B}_m(\tau ,\beta ) &=&\frac 1\pi \int_0^\infty B_m(\beta H)\sin
%(H\tau )dH.  \label{eq15}
%\end{eqnarray}
\end{multicols}
\widetext
\noindent\rule{20.5pc}{0.1mm}\rule{0.1mm}{1.5mm}\hfill

\begin{eqnarray}
a_m &=&\int_{-\infty }^\infty {\cal A}_m(\tau ,\beta )\exp \left\{
\sum_{m^{^{\prime }}=1/2}^Sa_{m^{\prime }}{\cal F}_1(m\tau )\right\} \cos
\left( \sum_{m^{\prime }=1/2}^Sb_{m^{\prime }}{\cal F}_2(m^{\prime }\tau
)\right) d\tau ;  \nonumber \\
b_m &=&\int_{-\infty }^\infty {\cal B}_m(\tau ,\beta )\exp \left\{
\sum_{m^{^{\prime }}=1/2}^Sa_{m^{\prime }}{\cal F}_1(m\tau )\right\} \sin
\left( \sum_{m^{\prime }=1/2}^Sb_{m^{\prime }}{\cal F}_2(m^{\prime }\tau
)\right) d\tau,   \label{eq14a} \\
{\cal  A}_m(\tau ,\beta ) &=&\frac 1\pi \int_0^\infty A_m(\beta H)\cos
(H\tau )dH;  \
{\cal B}_m(\tau ,\beta ) =\frac 1\pi \int_0^\infty B_m(\beta H)\sin
(H\tau )dH.
\end{eqnarray}

\hfill\rule[-1.5mm]{0.1mm}{1.5mm}\rule{20.5pc}{0.1mm}
\begin{multicols}{2}
\narrowtext
To obtain (\ref{eq14a}), we substite of (\ref{eq13a}) into Eq.(\ref{eq13}).

Solutions of the Eqs (\ref{eq14a}) allow to obtain magnetization of the
system ${\cal M}=g\mu M_1$, where $M_1\equiv -\left\langle \left\langle
S_Z\right\rangle \right\rangle =-\sum_{m=1/2}^Smb_m$ as well as other spin
averages
\begin{equation}
M_k\equiv \left\langle \left\langle S_Z^k\right\rangle \right\rangle =\left(
-1\right) ^k\sum_{m=1/2}^Sm^k \left\{
\begin{array}{c}
a_m\text{; for even }k, \\
b_m\text{; for odd }k.
\end{array}
\right.  \label{eq14b}
\end{equation}

The critical temperature $T_c$ is usually defined as a temperature, where
nonzero infinitesimal magnetization appears. In terms of Eqs (\ref{eq14a})
%Let we put $H_0=0$ and determine
%the critical temperature $T_c$ with
%respect to parity (''chetnost''') breaking of distribution function $f(H)$.
this means that we can put $b_m\to 0$ for $T\geq T_c$.
%because $B_m(H) $ are odd functions.
In this case, the parameters $a_m=a_m(\beta )
$ should be found from the equations
\begin{equation}
a_m=\int_{-\infty }^\infty {\cal A}_m(\tau ,\beta )\exp \left\{
\sum_{m^{^{\prime }}=1/2}^Sa_{m^{\prime }}{\cal F}_1(m^{\prime }\tau
)\right\} d\tau ;  \label{eq16}
\end{equation}
that stem from (\ref{eq14a}) for $b_m=0$.

Parameters $b_m$ can be obtained by linearization of the second equation (\ref{eq14a}):

\end{multicols}
\widetext
\noindent\rule{20.5pc}{0.1mm}\rule{0.1mm}{1.5mm}\hfill

\begin{equation}
b_m=\sum_{m^{\prime }=1/2}^SK_{m,m^{\prime }}b_{m^{\prime }}; \
%\label{eq17}
K_{m,m'}=\int_{-\infty }^\infty {\cal B}_m(\tau ,\beta ){\cal %
F}_2(m^{\prime }\tau )\exp \left\{ \sum_{m^{\prime \prime
}=1/2}^Sa_{m^{\prime \prime }}(\beta ){\cal F}_1(m^{\prime \prime }\tau
)\right\} d\tau .  \label{eq19}
\end{equation}

\hfill\rule[-1.5mm]{0.1mm}{1.5mm}\rule{20.5pc}{0.1mm}
\begin{multicols}{2}
\narrowtext
Thus, near $T_c$ parameters $b_m$ obey the system of homogeneous linear equations.
Well-known condition of existence of
non-trivial solution of this system gives following equation for $T_c$:
\begin{equation}
\det \left| {\bf K}-{\bf I}\right| =0,  \label{eq18}
\end{equation}
where ${\bf I}=\delta _{m,m'}$ is the identity matrix,
$m,m'=1/2,...,S$.

The Eqs (\ref{eq14a}) and (\ref{eq18}) are the main result of theoretical
background we develop. We start the analysis of these equations from the
case of $S=1/2$. The system (\ref{eq19}) reduces to single
equation with $a_{1/2}=A_{1/2}=1;{\cal A}_{1/2}\left( \tau ,\beta \right)
=\delta \left( \tau \right) $ and $B_{1/2}=\tanh \beta H/2$; ${\cal B}%
_{1/2}\left( \tau ,\beta \right) =(\beta \sinh \pi \tau /\beta )^{-1}$.

Subsequent calculations require the definition of spatial dependence
of $J(\vec r)$. Usually in the problems of carrier -induced
ferromagnetism in DMS, the RKKY interaction\cite{KittelBook} is
considered as an effective spin-spin exchange interaction that results in FM
ordering. Recently, the complex valence band structure of III-V and II-VI
semiconductors was discussed as a reason for magnetic anisotropy and some
enhancement of carrier induced magnetic interaction\cite{McDon01},\cite
{Dietl01}. To clarify the role of fluctuations, here we neglect
the effects of complex valence band structure, which can be also
incorporated in our theory.

In the case of simple one band carrier structure, the RKKY interaction reads
\begin{eqnarray}
J(r) &=&J_0\frac{x\cos x-\sin x}{x^4}, \ x=2k_Fr \label{mu1} \\
\ J_0 &=&\frac 1{4\pi ^3}\frac{J_{ci}^2\Omega _0^2m^{*}k_F^4}{\hbar ^2},
\end{eqnarray}
where $J_{ci}$ is a carrier-ion exchange constant, $\Omega _0$ is a unit cell volume,
$m^{*}$ is the density of states effective mass. The critical
temperature is determined now by the equation that follows from the Eq. (\ref
{eq19}) after the substitution of Eq. (\ref{mu1}) to integrals (\ref{eq13a}):
\begin{equation}
1=\frac{\theta \nu }2\int_{-\infty }^\infty \frac{\varphi _2\left( \xi
\right) }{\sinh \left( \theta \xi \right) }\exp \left( -\frac{\pi \nu }%
2\varphi _1\left( \xi \right) \right) d\xi .  \label{s19}
\end{equation}
Here we introduce the dimensionless parameters $\theta =\pi T_c/2J_0
$ and $\nu =n_i/6\pi ^2n_e$. The
functions $\varphi _1\left( \xi \right) $ and $\varphi _2\left( \xi \right) $
are related to integrals (\ref{eq13a}). In the case of homogeneous
magnetic ions distribution, $n_i({\vec r})=n_i=const,$%
\begin{eqnarray}
\varphi _1\left( \xi \right) &=&\int_0^\infty \left\{ 1-\cos \left( \xi \frac{%
x\cos x-\sin x}{x^4}\right) \right\} x^2dx  \label{s20a}\\
\varphi _2\left( \xi \right) &=&\int_0^\infty \sin \left( \xi \frac{x\cos
x-\sin x}{x^4}\right) x^2dx.  \label{s20b}
\end{eqnarray}

The MFA results (for spin $S=1/2$) can be recovered from Eqs (\ref
{s20a}) and (\ref{s20b}) by their expansion up to the linear terms
$\varphi _1\left( \xi \right) \to 0$;
$\varphi _2\left( \xi \right) \to \xi $, that
leads to the equation for critical temperature in the form
\[
\theta ^{MF}=\nu \pi ^2/4.
\]
Gaussian asymptotics of distribution function can be used to
analyze the influence of fluctuations on $T_c$ near MFA.
%the approximation of distribution function by Gaussian can
Latter asymptotics corresponds to the next term of expansion of the Eq. (%
\ref{s20a}),
\begin{equation}
\varphi _1\left( \xi \right) \to \pi \xi ^2/30, \ \varphi
_2\left( \xi \right) \to \xi .\label{ga}
\end{equation}

The numerical solution of the Eq. (\ref{s19}) with respect to (\ref{ga})
is presented in the Fig.1. Note the
qualitative difference in behavior of $\theta $ as a function of $n_e/n_i$
for MFA and Gaussian distribution function. While MFA predicts
the FM ordering at any relation between the electron and magnetic ion
concentrations, the fluctuations suppress ferromagnetism and make it
impossible if the concentration ratio $n_e/n_i=1/6\pi ^2\nu $ exceeds
critical value $(n_e/n_i)_c=1/6\pi ^2\nu _c$. We find
\begin{equation}
(\nu _c)_{\rm {Gauss}}=\pi /15 \approx 0.2094\label{crit_ga}
\end{equation}
as a solution of the Eq. (\ref{s19}) in the limit $\theta \to 0$
that corresponds to fluctuation restriction for carrier concentration
$((n_{ec})_{\rm {Gauss}}=5/2\pi ^3\approx 0.08n_i$.

\begin{figure}[th]
\vspace*{-5mm}
\centerline{\centerline{\psfig{figure=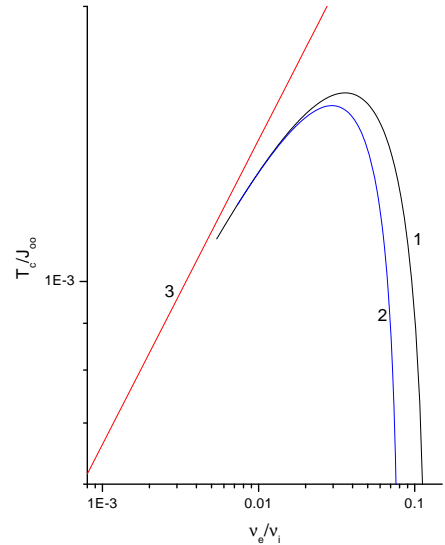,width=0.8\columnwidth}}}
%\vspace*{0.5mm}
\caption{Dimensionless phase transition temperature $T_c$ $/$ $J_{00}$ for spin 1/2 versus
ratio $\nu _e$ $/$ $\nu _i$. Here $J_{00}=\frac{(6\pi ^2)^{4/3}}{4\pi ^3}\frac{J_{ci}^2m^*\Omega _0^{2/3}}{\hbar^2}$,
 $\nu _{i,e}=n_{i,e}\Omega _0$, $\Omega _0$ is unit cell volume. Here we put $\nu _i=1$. 1- non-Gaussian distribution function, 2- Gaussian distribution function, 3 - MFA.}
%\label{diagram}
\end{figure}

We can also take into account the deviations of molecular field fluctuations
from their Gaussian asymptotics by numerical solution of the Eq. (\ref
{s19}).
%To simplify calculations the integrals (\ref{s20a}) and (%
%\ref{s20b}) have been approximated by the functions $\varphi _1(\xi )\approx
%\frac \pi {30}\xi ^2/\left( 1+0.05401\xi ^{1.36}\right) $ and $\varphi
%_2(\xi )\approx \xi /\left( 1+0.008625\xi ^{1.853}\right) $ with accuracy $%
%1\%$.
The results are also reported in the Fig.1. The critical concentration for this
case turns out to be
\begin{equation}
\nu _c \approx 0.1366, \ n_{ec}\approx 0.12n_i \label{crga}
\end{equation}
%It is seen that this value has the largest deviation from
The visible deviation of $T_c$ (\ref{s19}) from that for
Gaussian asymptotics is revealed at relatively narrow interval
$n_e/n_i$ near the critical value $(n_e/n_i)_c$. It can be shown that this property
holds for any other magnetic ion spin $S>1/2$. This permits to assert that the
molecular field fluctuations can be considered with sufficient accuracy
with Gaussian asymptotics for distribution function.

Gaussian asymptotics for $S>1/2$ corresponds to the expansion of Eq. (\ref
{eq13a}) in powers of $t$ up to the second order (it can be shown that this is valid for
sufficiently high spin concentration):
\begin{eqnarray}
{\cal F}_1(m\tau ) &=&-\sigma ^2m^2\tau ^2,  \nonumber \\
{\cal F}_2(m\tau ) &=&-\Delta Hm\tau   \label{eq33}
\end{eqnarray}
with
\begin{eqnarray}
\sigma ^2&=&\frac 12\int_Vn_i(\vec r)J^2(\vec r)d^3r,  \label{eq33a}\\
\Delta H&=&\int_Vn_i(\vec r)J(\vec r)d^3r.  \label{eq33b}
\end{eqnarray}
Approximation (\ref{eq33}) permits to find the explicit form for
Gaussian asymptotics of distribution
function (\ref{eq6}). This function depends on only two parameters (%
\ref{eq14b}) $M_1$ and $M_2$:
\begin{eqnarray}
f_G(H)&=&f_{M_1,M_2}(H)= \nonumber \\
&=&\frac 1{2\sigma \sqrt{\pi M_2}}\exp \left[ -\frac{%
\left( H-\Delta HM_1\right) ^2}{4\sigma ^2M_2}\right] .  \label{eq38}
\end{eqnarray}
Thus, instead the $(2S+1)/2$ equations (\ref{eq14a}) the Gaussian
asymptotics "generates" only two of them
\begin{eqnarray}
M_1 &=&\int_{-\infty }^\infty \left\langle S_Z\right\rangle
_{H_f}f_{M_1,M_2}(H_f)dH_f,\   \label{eq36} \\
M_2 &=&\int_{-\infty }^\infty \left\langle S_Z^2\right\rangle
_{H_f}f_{M_1,M_2}(H_f)dH_f.  \label{eq36a}
\end{eqnarray}
Note that the equations (\ref{eq36}) and (\ref{eq36a})
resemble replica-symmetric solution found for Ising (spin $1/2$)
spin glass (see \cite{Corenblit}). The reason we do not use replica formalism here
is that the explicit form of $J({\vec r})$ is important for DMS.
The MFA result immediately follows
from the Eq. (\ref{eq36}) in the limit of zero
fluctuations, $\sigma \to 0$, $f_{M_1,M_2}(H)\to \delta
(H-\Delta HM_1)$:
\begin{equation}
T_c^{MF}=\frac 13S(S+1)\int_Vn_i(\vec r)J(\vec r)d^3r.  \label{TcMF}
\end{equation}

The equation for $T_c$ can be obtained from Eq's (\ref{eq36}) and (\ref{eq36a})
as $M_1\to 0$.  Then Eq.(\ref{eq36}) can be transformed to following equation for
critical temperature
\begin{equation}
T_c=\frac{\Delta H}{\sqrt{\pi }}\int_{-\infty }^\infty m_S'\left(
\frac{2\sigma \sqrt{M_2}}{T_c}x\right) e^{-x^2}dx,  \label{eq45a}
\end{equation}
where $m_S\left( x\right) =SB_S(Sx)$ is non-normalized Brillouin function, $%
m_S^{\prime }\left( x\right) $ is its derivative. In the case of small
fluctuations, $2\sigma \sqrt{M_2}<<T$, Eq. (\ref{eq45a}) is reduced to MFA
result (\ref{TcMF}).

We can find analytically the first fluctuation correction to $T_c^{MF}$
due to Gaussian fluctuations of molecular field. The first term of expansion
of (\ref{eq45a}) in $\sigma /\Delta H$ yields

\begin{equation}
\frac{T_c}{T_c^{MF}}=1-\frac 65\left( 1+\frac 1{2S(S+1)}\right) \frac{\sigma
^2}{\Delta H^2}.  \label{eq45b}
\end{equation}
The physical meaning of Eq.(\ref{eq45b}) is the lowering of $T_c$ by the fluctuations.
This means that due to fluctuations, FM ordering occurs at lower temperature or that
the fluctuations suppress FM order at a given temperature.

To apply this result to the case under
consideration, the parameters (\ref{eq33a}) and (\ref{eq33b}) have been evaluated for
RKKY interaction (\ref{mu1}) and constant spins concentration $n_i(\vec
r)=n_i=const$:
\begin{eqnarray}
\Delta H &=&\frac{J_0}{12\pi }\frac{n_i}{n_e},  \label{eq34} \\
\sigma  &=&\frac{J_0}{6\sqrt{10}}\sqrt{\frac{n_i}{n_e}}.  \nonumber
\end{eqnarray}
This gives
\begin{equation}
\left(\frac{T_c}{T_c^{MF}}\right)_{\rm {RKKY}}=%
1-\frac{12\pi ^2}{25}\left( 1+\frac 1{2S(S+1)}\right) \frac{n_e}{n_i}.
\label{eq46}
\end{equation}
For the important case of the magnetic ions $Mn^{2+}$ with $S=5/2$ the Eq. (%
\ref{eq46}) yields $1-5.0(n_e/n_i)$.

\begin{figure}[th]
\vspace*{-5mm}
\centerline{\centerline{\psfig{figure=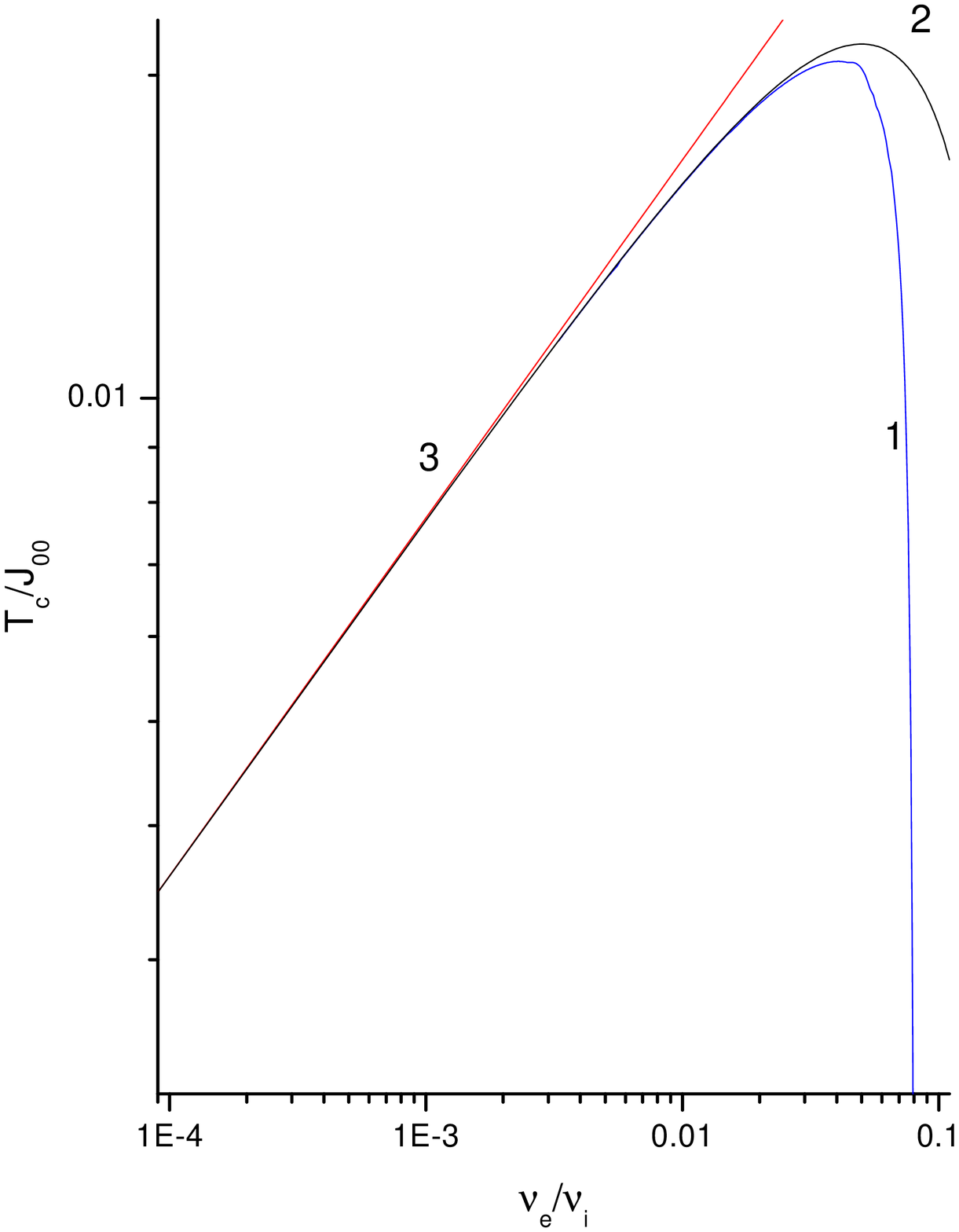,width=0.8\columnwidth}}}
%\vspace*{0.5mm}
\caption{Dimensionless phase transition temperature $T_c/J_{00}$ for spin 5/2 versus
ratio $\nu _e/\nu _i$ (also $\nu _i=1$). The notations are similar to those for spin 1/2.
1- Gaussian distribution function, 2- first fluctuation correction to MFA, 3 - MFA.}
%\label{diagram}
\end{figure}

In Fig.2, we reproduce the electron concentration dependence of
$T_c$ calculated in MFA, with Gaussian distribution function (\ref{eq45a})
and with approximation (\ref{eq46}) for $S=5/2$.
The dependence $T_c=T_c(n_e/n_i)$ presented in the figure reveals
also the presence of critical ratio $(n_e/n_i)_c$. The
equation for the $(n_e/n_i)_c$ can be found as $T\to 0$ in
Eq.(\ref{eq45a}). Equation we obtain ($(\Delta H)^2=\pi
\sigma ^2$) is independent of spin, so account for the Eqs (\ref{eq34})
reproduces the result we discussed for spin $S=1/2$: $\nu _c=\pi /15$,
or $(n_e/n_i)_c=5/2\pi ^3\simeq 0.08$.

In conclusion, here we present a theory of fluctuated molecular field for the
spin systems with given dependence $J(\vec r)$. We derive the equations
for order parameters and FM phase transition temperature
for Ising model with indirect spin-spin RKKY interaction in DMS.
Our analysis show, that fluctuations of transverse spin components (which are present in Heisenberg
model) do not reveal new qualitative properties in the problem of FM ordering in DMS.

Naturally, the RKKY interaction is not unique spin-spin interaction in DMS.
There is additional background interaction independent of free carrier
concentration. In II-VI DMS this AFM exchange interaction is dominant
for close magnetic pairs. An intuitive approach to this problem was
proposed in Ref\cite{Cibert}, where MFA was applied to ensemble of magnetic
ion spins with excluded AFM exchange pairs. Our formalism can
naturally incorporate this interaction by adding correspondent terms to Hamiltonian (%
\ref{mu0}).

Actually, the parameter of theory that present the fluctuations of RKKY
interaction in bulk DMS is a cube of ratio of Fermi wave vector length to
mean inter-ion separation. We can expect that similar parameter
will be "responsible" for fluctuations in 2D systems. This mean that
critical temperature would depend on carrier concentration in accordance with
MFA prediction in 2D case.

The detailed analysis of all aforementioned factors will be presented elsewhere.

\acknowledgments

This work was partially supported by the Grant No. 09244106 of the Ministry
of Education and Science of Japan.

\end{multicols}

\end{document}